\documentclass[aps,prc,twocolumn,showpacs,superscriptaddress]{revtex4-1}
\usepackage{float}
\usepackage{natbib}
\usepackage{graphicx}
\usepackage[usenames,dvipsnames]{xcolor}
\usepackage{amsfonts}
\usepackage{amsmath,amssymb}
\usepackage{lineno,soul}

\newcommand{\lwk}{{{\rm low}\mbox{-}k}}
\newcommand{\bb}{$\beta$}
\newcommand{\vlwk}{$V_{{\rm low}\mbox{-}k}$}
\newcommand{\thetaeff}{$\Theta_{\rm eff}$}

\newcommand{\zbb}{$0\nu\beta\beta$}
\newcommand{\dbb}{$2\nu\beta\beta$}
\newcommand{\heff}{$H_{\rm eff}$}

\newcommand{\qbox}{$\hat{Q}$~box}
\newcommand{\tbox}{$\hat{\Theta}$~box}
\newcommand{\nme}{$M^{0\nu}$}
\newcommand{\nmes}{$M^{0\nu}$s}

\newcommand{\logft}{log$ft$}
\newcommand{\logfts}{log$ft$s}
\newcommand{\In}{$^{115}$In}
\newcommand{\Sn}{$^{115}$Sn}
\newcommand{\Cd}{$^{113}$Cd}
\newcommand{\Inn}{$^{113}$In}
\newcommand{\Nb}{$^{94}$Nb}
\newcommand{\Mo}{$^{94}$Mo}
\newcommand{\Tc}{$^{99}$Tc}
\newcommand{\Ru}{$^{99}$Ru}

\begin{document}

\title{Study of forbidden $\beta$ decays within the realistic shell model}

\author{G. De Gregorio}
\affiliation{Dipartimento di Matematica e Fisica, Universit\`a degli
  Studi della Campania ``Luigi Vanvitelli'', viale Abramo Lincoln 5 -
  I-81100 Caserta, Italy}
\affiliation{Istituto Nazionale di Fisica Nucleare, \\ 
Complesso Universitario di Monte  S. Angelo, Via Cintia - I-80126 Napoli, Italy}
\author{R. Mancino}
\affiliation{Institute of Particle and Nuclear Physics, Faculty of Mathematics and Physics,  
Charles University, V Hole\v sovi\v ck\'ach 2, 180 00 Prague, Czech Republic } 
\affiliation{Institut f{\"u}r Kernphysik (Theoriezentrum), Fachbereich
  Physik, Technische Universit{\"a}t Darmstadt, \\
Schlossgartenstrasse 2 - 64298 Darmstadt, Germany}
\affiliation{GSI Helmholtzzentrum f{\"u}r Schwerionenforschung,
  Planckstrasse 1, 64291 Darmstadt, Germany}
\author{L. Coraggio}
\affiliation{Dipartimento di Matematica e Fisica, Universit\`a degli
  Studi della Campania ``Luigi Vanvitelli'', viale Abramo Lincoln 5 -
  I-81100 Caserta, Italy}
\affiliation{Istituto Nazionale di Fisica Nucleare, \\ 
Complesso Universitario di Monte  S. Angelo, Via Cintia - I-80126 Napoli, Italy}
\author{N. Itaco}
\affiliation{Dipartimento di Matematica e Fisica, Universit\`a degli
  Studi della Campania ``Luigi Vanvitelli'', viale Abramo Lincoln 5 -
  I-81100 Caserta, Italy}
\affiliation{Istituto Nazionale di Fisica Nucleare, \\ 
Complesso Universitario di Monte  S. Angelo, Via Cintia - I-80126 Napoli, Italy}

\begin{abstract}
For the first time, half-lives and energy spectra of forbidden \bb~
decays are calculated within the realistic shell model.
Namely, we approach this issue starting from a realistic
nucleon-nucleon potential and deriving effective Hamiltonians
and decay operators.
Our goal is to explore the sensitivity of the shape of calculated
energy spectra to the renormalization of forbidden \bb-decay
operators, an operation that allows to take into account those
configurations that are not explicitly included in the chosen model
space.
The region that has been considered for this investigation are nuclei
outside the $^{78}$Ni core, more precisely we have studied the
second-forbidden \bb~decays of \Nb~ and \Tc, and fourth-forbidden
$\beta$ decays of \Cd~ and \In, that are currently of a renewed
experimental interest in terms of novel spectroscopic techniques.
Our results evidence that the introduction of a renormalized \bb-decay
operator leads to a marked improvement of the reproduction of
experimental half-lives.
As regards the spectra of both second-forbidden  and fourth-forbidden
decays, we have found that their calculated shapes are in good
agreement with the observed ones, even if scarcely responsive to the
renormalization of the decay operator.
We carry out also a detailed inspection of the different components of
the calculated spectra for a deeper insight about their role in
reproducing the experimental shapes.
\end{abstract}


\maketitle

\section{Introduction}
\label{intro}
The understanding of the renormalization mechanisms of electroweak
currents is nowadays a cornerstone of the nuclear structure research.
The attention to this issue is motivated by the need of calculating
reliable nuclear matrix elements \nme~for the \zbb~decay, and relating
the inverse half-life $\left[ T^{0\nu}_{1/2}\right]^{-1}$ of such a
rare process to the neutrino effective mass.

As a matter of fact, the accurate calculation of the wave functions of
parent and grand-daughter nuclei does not ensure a trustable \nme,
since most nuclear models are based on the reduction of the dimension
of the Hilbert space where the nuclear Hamiltonian is defined.
Then, a sound knowledge of the renormalization of the electroweak
currents, to account for the configurations that are not explicitly
included in the components of the nuclear wave function, is crucial to
enhance the predictivity of the calculated \nmes.

The ability of nuclear models to reproduce \bb-decay observables is,
consequently, the better way to validate both wave functions and
renormalization procedures, and such an issue is connected to the
so-called  ``quenching puzzle" of the axial coupling constant, namely
the need by most nuclear structure models to resort to a reduction of
$g_A$ to reproduce the observables directly linked to Gamow-Teller
(GT) transitions \cite{Martinez-Pinedo96,Barea15,Suhonen19}.
However, this is an empirical procedure, and it cannot be generalized
to any \bb-decay operator that depends on the value of the axial
coupling constant.

The realistic shell model (RSM) provides a consistent approach to
derive effective Hamiltonians and decay operators, the only parameter
that is involved being the nuclear force one starts from.
In such a framework, single-particle (SP) energies and two-body matrix
elements (TBMEs) of the effective shell-model Hamiltonian \heff, as
well as every matrix element of decay operators, are derived from a realistic
free nucleon-nucleon ($NN$) potential $V_{NN}$ by way of the many-body
theory \cite{Kuo90,Suzuki95}.
The bare matrix elements of the $NN$ potential, and of any transition
operator, are renormalized with respect to the truncation of the full
Hilbert space into the reduced shell-model (SM) model space, to take
into account the neglected degrees of freedom without resorting to
any empirical parameter \cite{Coraggio20c}.
In other words, this approach does not apply effective charges to
calculate electromagnetic transition strengths, and empirical
quenching of $g_A$ to reproduce the \bb-decay matrix elements.

We have successfully employed RSM to study the \dbb-decay
of $^{48}$Ca, $^{76}$Ge, $^{82}$Se,$^{100}$Mo, $^{130}$Te, and
$^{136}$Xe \cite{Coraggio17a,Coraggio19a,Coraggio22a,Coraggio24a}, and
then extended it to predict the nuclear matrix elements of their
\zbb-decay \cite{Coraggio20a,Coraggio22a}.
Now, in order to validate the RSM in predicting \bb-decay observables,
in the present work we investigate the sensitivity to the
renormalization of SM forbidden \bb-decay operators describing the
energy spectra of the emitted electrons.

To this end, we have considered the second-forbidden \bb-decays of
\Nb~ and \Tc~ into \Mo~ and \Ru, as well as the fourth-forbidden
\bb-decays of \Cd~ and \In~ into \Inn~ and \Sn,
respectively, and compared their calculated \logfts~ and \bb-decay
energy spectra as obtained with the bare and the renormalized decay
operators, as well as with the available data.

The motivations of such a choice are twofold: first, these decays have
been already the subject of a few works with a similar goal, where the
effective operators was obtained tuning the quenching factor $q$ of
$g_A$, and the focus was spotted on the dependence of the shape of
energy spectra on the value of $q$
\cite{Mustonen06,Haaranen16,Haaranen17,Kostensalo21,Kostensalo23}.
Second, novel experimental techniques have triggered new measurements
of the energy spectra of \bb~decays in the region of the \zbb~decay of
$^{100}$Mo.
Among them, we mention the COBRA demonstrator \cite{Zuber01}, that has
been developed for double-$\beta$ decay experiments, and also adopted
to study the spectra of \bb~decays of
\Cd~\cite{Goessling05,Boden20,Kostensalo21}; the ACCESS project that
aims to perform precision measurements of forbidden \bb-decays using
cryogenic calorimeters \cite{Pagnanini23}.
Another new experimental project is ASPECT-BET (An SDD-SPECTrometer
for BETa decay studies), that has developed a new detection strategy
based on silicon drift detectors (SDD), and it should be able to
perform high-precision, high-accuracy measurements of the energy
spectra of \bb~ decays at room temperature \cite{Biassoni23}.

This paper is organized as follows.
\noindent
First, in Sec. \ref{outline} we sketch out briefly the derivation of
the effective SM Hamiltonian and decay operators, as well as the basic
theory of the \bb-decay and the structure of the second- and
fourth-forbidden \bb-decay matrix elements.

The effective shell-model Hamiltonian and decay operators have been
derived within a model space that is spanned by four
$0f_{5/2},1p_{3/2},1p_{1/2},0g_{9/2}$ proton orbitals and five
$0g_{7/2},1d_{5/2},1d_{3/2},2s_{1/2},0h_{11/2}$ neutron orbitals
outside $^{78}$Ni core starting from the high-precision CD-Bonn $NN$
potential \cite{Machleidt01b}, whose repulsive high-momentum
components are renormalized using the \vlwk~procedure
\cite{Bogner02}.
This is the same framework we have employed in our previous study of
the double-\bb~decay of $^{100}$Mo \cite{Coraggio22a}.

The results of the shell-model calculations are discussed and compared
with the available experimental data in Sec. \ref{results}. 
There, we check first our nuclear wave functions by comparing the
calculated low-energy spectra and $E2$ transition strengths of parent
and daughter nuclei, which are involved in the decays under
consideration, with their experimental counterparts.
Then, we report the results of our theoretical \logfts~ and energy
spectra of the emitted electron and size up them to the available
data.
We complete our analysis with a detailed analysis of the different
components of the spectra, and how the interplay among their
contributions plays an important role in reproducing data.

In the last section (Sec. \ref{conclusions}), we summarize the
conclusions of this study, as well as the outlook of our current
research project.

\section{Outline of the theory}\label{outline}
\subsection{The effective SM Hamiltonian}\label{effham}
The procedure of the derivation of the effective SM Hamiltonian is the
same as the one followed in Ref. \cite{Coraggio22a}.
First, we consider the high-precision CD-Bonn $NN$ potential
\cite{Machleidt01b}, then the non-perturbative repulsive high-momentum
components are integrated out, by way of the \vlwk~unitary
transformation \cite{Bogner02,Coraggio09a}, that provides a smooth
potential preserving all the two-nucleon observables calculated with
the CD-Bonn one.

The \vlwk~matrix elements are chosen as the interaction vertices of a
perturbative expansion of \heff, as well as those of the Coulomb
interaction between protons, and a detailed description of the
many-body perturbation theory approach to the nuclear \heff~can be
found in Refs. \cite{Hjorth95,Coraggio12a,Coraggio20c}, so here we
only sketch briefly the steps that have been followed to obtain
\heff.

The starting point is the full nuclear Hamiltonian $H$ for $A$ interacting
nucleons, which, according to the SM ansatz, is divided into the
one-body term $H_0$, whose eigenvectors set up the SM basis, and the
residual interaction $H_1$, using the harmonic-oscillator (HO)
auxiliary potential $U$:

\begin{eqnarray}
 H &= & T + V_\lwk = (T+U)+(V_\lwk-U)= \nonumber\\
~& = &H_{0}+H_{1}~.\label{smham}
\end{eqnarray}

The eigenvalue problem of $H$ for a many-body system,
in an infinite basis of eigenvectors of $H_0$, cannot be solved, then
an effective Hamiltonian is derived, which is defined in the truncated
model space spanned by four proton -- $0f_{5/2}, 1p_{3/2}, 1p_{1/2},
0g_{9/2}$ -- and five neutron orbitals -- $0g_{7/2}, 1d_{5/2},
1d_{3/2}, 2s_{1/2}, 0h_{11/2}$ -- outside $^{78}$Ni core.

The effective Hamiltonian is derived by way of the time-dependent
perturbation theory, through the Kuo-Lee-Ratcliff folded-diagram
expansion in terms of the $\hat{Q}$-box vertex function
\cite{Kuo90,Hjorth95,Coraggio12a}:

\begin{equation}
H^{\rm eff}_1 (\omega) = \hat{Q}(\epsilon_0) - P H_1 Q \frac{1}{\epsilon_0
  - Q H Q} \omega H^{\rm eff}_1 (\omega) ~,\label{eqfinal}
\end{equation}
\noindent
where $\omega$ is the wave operator decoupling the model space $P$ and
its complement $Q$, and $\epsilon_0$ is the eigenvalue of the unperturbed
degenerate Hamiltonian $H_0$.

The \qbox~is defined as
\begin{equation}
\hat{Q} (\epsilon) = P H_1 P + P H_1 Q \frac{1}{\epsilon - Q H Q} Q
H_1 P ~, \label{qbox}
\end{equation}
\noindent
and $\epsilon$ is an energy parameter called ``starting energy''.

Since the exact calculation of the \qbox~is not possible, then the
term $1/(\epsilon - Q H Q)$ is expanded as a power series

\begin{equation}
\frac{1}{\epsilon - Q H Q} = \sum_{n=0}^{\infty} \frac{1}{\epsilon -Q
  H_0 Q} \left( \frac{Q H_1 Q}{\epsilon -Q H_0 Q} \right)^{n} ~,
\end{equation}

\noindent
namely we perform an expansion of the \qbox~up to the third order in
perturbation theory \cite{Coraggio20c}.

We would point out that, as in previous works
\cite{Coraggio17a,Coraggio19a, Coraggio20a,Coraggio22a}, we include a
number of intermediate states in the perturbative expansion of the
shell-model effective Hamiltonian and decay operators, whose maximum
allowed excitation energy -- expressed in terms of the number of
oscillator quanta $N_{\rm max}$ \cite{Coraggio12a} -- is $N_{\rm
  max}$=16.
This set of intermediate states is sufficient to obtain convergent
values of the single-particle energies, TBMEs, and matrix elements of
the decay operators, as it has been shown in
Refs. \cite{Coraggio18,Coraggio20a}.

As a matter of fact, the calculation of the \qbox~is the start to
solve the non-linear matrix equation (\ref{eqfinal}) and obtain
\heff~by way of iterative techniques such as the Kuo-Krenciglowa and
Lee-Suzuki ones \cite{Krenciglowa74,Suzuki80}, or graphical
non-iterative methods \cite{Suzuki11}.

It should be pointed out that, since the nuclei that are involved in
the decay processes under investigation are characterized by a large
number of valence nucleons, we have included contributions from
induced three-body forces in the calculation of the \qbox, that
involve also three valence nucleons.

Since the SM code we have employed for our calculations \cite{KSHELL}
cannot diagonalize a three-body \heff, we have performed a
normal-ordering decomposition of the three-body induced-force
contributions arising at second order in perturbation theory, and
retained only the two-body term that is density-dependent from the
number of valence nucleons.
This procedure is presented in details in
Refs. \cite{Coraggio20c,Coraggio20e}, together with a discussion about
the contribution of such terms to the eigenvalues of the SM Hamiltonian.

The SM parameters, namely the SP energies and the TBMEs of the
residual interaction, are reported in the Supplemental Material
\cite{supplemental2024}.

\subsection{ \bb-decay theory}
\label{operators}
The  theory of \bb-decay is here briefly outlined. 
More details can be found in Refs. \cite{Behrens71,Dag20}. 

In the following we focus on the \bb$^-$-decay, moreover, we use
natural units ( $\hbar=c=m_e=1$).

The total half-life of the \bb~decay is expressed in terms of the
$k$-th partial decay half-life $t^k_{1/2}$ as follows:
\begin{equation}
\frac{1}{T_{1/2}}=\sum_k \frac{1}{t^k_{1/2}}~.
\end{equation}

On the other hand, the partial half-life $t^k_{1/2}$ is related to the
dimensionless integrated shape function $\tilde{C}$ by way of the
relation:
\begin{equation}
\label{halflife} 
t^k_{1/2}=\frac{\kappa}{\tilde{C}}~,
\end{equation}
\noindent
where $\kappa=6144 \pm 2$ $s$ \cite{Hardy09}.

For a given $k$-th final state, the integrated shape function
$\tilde{C}$ -- whose integrand defines the \bb-decay energy spectrum
-- is written as
\begin{equation}
\label{intCeq}
\tilde{C}=\int^{w_0}_{1} C(w_e)p_ew_e(w_0-w_e)^2F(Z,w_e) dw_e ~.
\end{equation}

The quantities on the right-hand side of the above definition are
listed as:
\begin{itemize}
\item [a)] $Z$ is the atomic number of the daughter nucleus, $w_e$
  the adimensional energy of the emitted electron, $w_0$ the endpoint
  energy -- namely the maximum electron energy for a given
  transition--, and $p_e$ the electron momentum.
\item [b)] The function $F(Z,w_e)$ is the Fermi function which is
  factorized in terms of two functions $F_0$ and $L_0$:
\begin{equation}
F(Z, w_e) = F_0 (Z, w_e) L_0(Z, w_e) ~,
\end{equation}
\noindent
where $F_0$ defines the effects of the Coulomb interaction between the
electron and the daughter nucleus, and $L_0$  accounts for the
electromagnetic finite-size effect, whose explicit expressions can be
found in Ref. \cite{Behrens71}.
\item [c)] $C(w_e)$ is the so-called nuclear shape function, which
  depends on the nuclear matrix elements (NMEs). For allowed $\beta$
  transitions, it corresponds to the GT reduced transition probability,
  and in such a case does not depend on the electron energy.
\end{itemize}

For $n$-forbidden transitions, $C(w_e)$ depends on the electron
energy, and is expressed as
\begin{eqnarray}
\label{CW}
    C(w_e) =\sum_K \sum_{k_e, k_\nu} \lambda_{k_e} \biggl[
  M_K^2(k_e,k_\nu) + m_K^2(k_e,k_\nu) - \nonumber \\ 
\frac{2 \gamma_{k_e}}{k_e w_e}M_K(k_e,k_\nu)m_K(k_e,k_\nu) \biggr]~,
\end{eqnarray}
\noindent
where $K$ is the tensor rank of the forbidden \bb-decay operators
involved in the decay.
$K$ can range from 0 to 2 for $n=1$, and from $n$ to $n+1$ for $n>1$. 
The quantities $k_e$ and $k_\nu$ are the positive integers emerging
from the partial wave expansion of the leptonic wave functions. 
The latter, for a given value of $K$, must satisfy either
$k_e+k_\nu=K+1$ or $k_e+k_\nu=K+2$.

The auxiliary quantities $\gamma_{k_e}$ and $\lambda_{k_e}$ are defined as 
\begin{equation}
\gamma_{k_e}=\sqrt{k_e^2-(\alpha Z)^2},\quad
\lambda_{k_e}=\frac{F_{k_e-1}(Z,w_e)}{F_0(Z,w_e)}~,
\end{equation}
\noindent
where $F_{k_e-1}(Z,w_e)$ is the so-called generalized Fermi function (see
Ref. \cite{Behrens71} for its explicit expression).
The quantities $M_K$ and $m_K$ are complicated combinations of some
kinematic factors and coefficients $F^{(N)}_{KLS}(k_e,m,n,\rho)$, the
latter being functions of the orbital, spin, and total rank of the
transition operators $L$, $S$, and $K$, respectively, and the integers
$m$, $n$ and $\rho$ depending on the nuclear charge distribution which
accounts for the influence of the nuclear charge on the electron
\cite{Behrens71,Dag20}.
The index $N$ labels the order of the expansion in powers of $qR$
(where $R$ is the nuclear radius and $q$ is the momentum transfer
$q=\mid p_e+p_\nu\mid$) of the nuclear form factor $F_{KLS}$, which is
defined by the following expression
\begin{equation}
\label{Fexp}
F_{KLS}(q^2)=\sum_N \frac{(-1)^N (qR)^{2N} (2L+1)!!}
{(2N)!!(2L+2N+1)!!}F^{(N)}_{KLS}~.
\end{equation}

If we adopt the impulse approximation ($qR\ll1$) to derive the
formalism of forbidden $\beta$-decay transitions, and also assume that
bound nucleons interact weakly as free nucleons, then we may neglect
the effect of any many-body current.
Within such an approximation, it has been shown in
Refs. \cite{Behrens71,Dag20} that the form factor coefficients
$^{V/A}F^{(N)}_{KLS}(k_e, m, n ,\rho)$ can be related to the NMEs
$^{V/A}M^{N}_{KLS}(k_e, m, n ,\rho)$ by a phase factor:
\begin{eqnarray}
&^{V/A}F^{(N)}_{KLS}(k_e, m, n ,\rho)=(-1)^{K-L}~
  ^{V/A}M^{N}_{KLS}(k_e, m, n ,\rho) ~,\nonumber \\
\end{eqnarray}
\noindent
where the label $V/A$ indicates the separation of the coefficients
$F^{(N)}_{KLS}(k_e, m, n ,\rho)$ according to the axial and vector
components of the decay operator.

In a shell-model calculation, the NMEs can be expressed in terms of
the single-particle matrix elements of the one-body decay operator
(SPMEs) and the one-body transition densities (OBTDs), that can be
obtained from the shell-model wave functions, through the expression
\begin{widetext}
\begin{equation}
^{V/A}M^{(N)}_{KLS}(k_e, m, n ,\rho)={\frac{1}{\hat{J_i}}}
\sum_{\pi,\nu} ~^{V/A}m^{(N)}_{KLS}(\pi,\nu) (k_e, m, n ,\rho) \times
{\rm OBTD}(\Psi_f,\Psi_i,\pi,\nu,K) ~,
\end{equation}
\end{widetext}
\noindent
where $\hat{J_i}=\sqrt{(2J_i+1)}$, and $J_i$ is the angular momentum
of the initial state of the parent nucleus.
The OBTDs are defined as

\begin{equation}
{\rm OBTD}(\Psi_f,\Psi_i,\pi,\nu,K)=\frac{\langle \Psi_f\mid\mid[
  a^\dagger_{\pi} \otimes \tilde{a}_{\nu}]^K\mid\mid
  \Psi_i\rangle}{\hat{K}}~,
\end{equation}

\noindent
where $\Psi_i$ and $\Psi_f$ are the wave function of the initial and
final state, respectively, $a^\dagger_{\pi}$ is the particle creation
operator, and $ \tilde{a}_{\nu}=(-1)^{j_\nu+m_\nu}a_{\nu-m_\nu}$ is
the tensor spherical form of the particle annihilation operator
($a_{\nu}$).
The indices $\pi$ and $\nu$ label the proton ($\pi$) and neutron
($\nu$) single-particle states, and the symbol $\otimes$ denotes the
angular-momentum coupling.

The SPMEs correspond to the following matrix elements:
\begin{widetext}
  \begin{eqnarray}
  \label{ME}
^{V}m^{(N)}_{KLS}(k_e, m, n ,\rho)=\langle \phi_{\kappa_\pi\mu}
  \mid\mid  \left(\frac{r}{R}\right)^{L+2N}\mathcal{I}(k_e, m, n
  ,\rho,r)T_{KLS}\mid\mid\phi_{\kappa_\nu\mu}\rangle~, \\
\label{ME2}
^{A}m^{(N)}_{KLS} (k_e, m, n ,\rho)=\langle
  \phi_{\kappa_\pi\mu}\mid\mid  \left(\frac{r}{R}\right)^{L+2N}
  \mathcal{I}(k_e, m, n ,\rho,r)\gamma_5 T_{KLS}
  \mid\mid\phi_{\kappa_\nu\mu}\rangle ~.
  \end{eqnarray}
\end{widetext}

Now, it is worth to list and specify the quantities which appear in
the Eqs. (\ref{ME},\ref{ME2}) for the matrix elements of the vector
and axial components of the $\beta$-decay operator:
\begin{itemize}
\item The functions $\mathcal{I}(k_e, m, n ,\rho,r)$ are the so-called
  Coulomb factors whose explicit expression can be found in
  Ref. \cite{Behrens71}.
\item The operator $T_{KLS}$ is the transition operator defined as
\begin{equation}
T^{M}_{KLS}=\left\{\begin{matrix}
Y_{LM}\delta_{KL} & S=0,\\ 
(-1)^{L-K+1} \gamma_5 [Y_L\otimes\sigma]_{KM} & ~S=1 ~.
\end{matrix}\right.
\end{equation}

\item The single-particle relativistic wave functions
  $\phi_{\kappa\mu}$ are the eigenfunctions of the operators $J_z$ and
  $\mathcal{K} = \beta (\sigma \cdot L+\mathbb{I})$, and are labeled
  by their eigenvalues $\mu$ and $\kappa$:
\begin{eqnarray}
J_z\phi_{\kappa \mu} & = & \mu\phi_{\kappa \mu}~, \nonumber\\
\mathcal{K}\phi_{\kappa \mu} &= &\beta(\sigma \cdot
L+\mathbb{I})\phi_{\kappa \mu}=\kappa\phi_{\kappa \mu}~. \nonumber
\end{eqnarray} 

It can be shown that the eigenvalue $\kappa$ is related to the total
and orbital angular momenta $j$ and $l$ through the relation
\begin{equation}
\kappa=\left\{\begin{matrix}
j+\frac{1}{2} & \textrm{for } l=j+\frac{1}{2} \\ 
-(j+\frac{1}{2}) & \textrm{for } l=j-\frac{1}{2}~.
\end{matrix}\right.
\end{equation}
\end{itemize}

In the Condon-Shortley (CS) phase convention, the $\phi_{\kappa\mu}$
functions are defined as
\begin{equation}
\label{spinor}
\phi_{\kappa \mu}=\binom{-i
  f_{\kappa}(r)\chi_{-\kappa\mu}}{g_{\kappa}(r)\chi_{\kappa\mu}}~,\\
~
\end{equation}
\noindent
where $\chi_{\kappa\mu}=[Y_l(\hat{r})\otimes \chi]^{j\mu}$, and the
radial functions $f_\kappa(r)$ and $g_\kappa(r)$ are the solutions of
the radial Dirac equations, and they are usually indicated as the {\it
  small} and {\it large} components, respectively:
\begin{eqnarray}
\frac{dg_\kappa(r)}{dr}+\frac{\kappa+1}{r}g_\kappa(r)-(E+M-V(r))f_\kappa(r)=0~,
  \label{dirac1}\\
\frac{df_\kappa(r)}{dr}-\frac{\kappa-1}{r}f_\kappa(r)+(E-M-V(r))g_\kappa(r)=0~. \label{dirac2}
\end{eqnarray}

The matrix elements of Eqs. (\ref{ME},\ref{ME2}) can be grouped into
two different types.

The first one contains the product of both small components of the
initial and final wave functions, as well as the product of both large
components of the same wave functions, which are the solutions of the
coupled equations (\ref{dirac1},\ref{dirac2}), and usually they are
dubbed in literature as the {\it non-relativistic} matrix elements
\cite{Behrens71,Behrens93,Zhi13,Dag20}.
Their explicit expression is:
\begin{widetext}
\begin{eqnarray} 
\label{SPME}
&^{V}m^{(N)}_{KK0}(\pi,\nu)(k_e,m,n,\rho) =  \sqrt{2}
 g_V \biggl[ G_{KK0}(\kappa_\pi,\kappa_\nu)  \int_0^\infty  g_{\pi}(r,\kappa_p) \left( \frac{r}{R}\right)^{K+2N}  \mathcal{I}(k_e,m,n,\rho,r) g_{\nu}(r,\kappa_\nu) r^2 dr\nonumber \\
 &+ G_{KK0}(-\kappa_\pi,-\kappa_\nu) \int_0^\infty f_{\pi}(r,\kappa_\pi) \left( \frac{r}{R}\right)^{K+2N} \mathcal{I}(k_e,m,n,\rho,r) f_{\nu}(r,\kappa_\nu) r^2 dr \biggr]~,\\
 &^{A}m^{(N)}_{KL1}(\pi,\nu)(k_e,m,n,\rho) =   \text{sign}(K - L + \frac{1}{2}) \sqrt{2} g_A\biggl[ G_{KK0}(\kappa_\pi,\kappa_\nu) \int_0^\infty  g_{\pi}(r,\kappa_\pi)\left( \frac{r}{R}\right)^{L+2N}  \mathcal{I}(k_e,m,n,\rho,r) g_{\nu}(r,\kappa_\nu) r^2 dr \nonumber \\ 
&+ G_{KK0}(-\kappa_\pi,-\kappa_\nu) \int_0^\infty f_{i_\pi}(r,\kappa_p) \left( \frac{r}{R}\right)^{L+2N} \mathcal{I}(k_e,m,n,\rho,r) f_{\nu}(r,\kappa_\nu) r^2 dr \biggr]~.
\end{eqnarray}
\end{widetext}

The second group contains the product of the small and large
components of the initial and final wave functions that are the
solutions of Eqs. (\ref{dirac1},\ref{dirac2}), and in such a case they
are dubbed as the {\it relativistic} matrix elements.
Now, their expression is:
\begin{widetext}
\begin{eqnarray}
\label{rSPME}
&^{V}m^{(N)}_{KL1}(\pi,\nu)(k_e,m,n,\rho) =  \text{sign}(K - L + \frac{1}{2}) \sqrt2
 g_V\biggl[ G_{KL1}(\kappa_\pi,-\kappa_\nu) \int_0^\infty  g_\pi(r,\kappa_\pi) \left( \frac{r}{R}\right)^{L+2N}  \mathcal{I}(k_e,m,n,\rho,r) f_\nu(r,\kappa_\nu) r^2 dr\nonumber \\
 &- G_{KK0}(-\kappa_\pi, \kappa_\nu) \int_0^\infty f_{\pi}(r,\kappa_\pi)\left( \frac{r}{R}\right)^{L+2N} \mathcal{I}(k_e,m,n,\rho,r) g_{\nu}(r,\kappa_\nu) r^2 dr \biggr]~,\\
&^{A}m^{(N)}_{KK0}(\pi,\nu)(k_e,m,n,\rho) = \sqrt2
  g_A\biggl[ G_{KK0}(\kappa_\pi,-\kappa_\nu) \int_0^\infty  g_{\pi}(r,\kappa_\pi) \left( \frac{r}{R}\right)^{K+2N}  \mathcal{I}(k_e,m,n,\rho,r) f_{\nu}(r,\kappa_\nu) r^2 dr \nonumber \\
  &- G_{KK0}(-\kappa_\pi,\kappa_\nu) \int_0^\infty f_{\pi}(r,\kappa_\pi) \left( \frac{r}{R}\right)^{K+2N} \mathcal{I}(k_e,m,n,\rho,r) g_{\nu}(r,\kappa_\nu) r^2 dr \biggr]~.
\end{eqnarray}
\end{widetext}

It is worth to note that in the above equations we have introduced the quantity
\begin{eqnarray}
G_{KLS}(\kappa_\pi,\kappa_\nu) = &(-1)^{j_\pi-j_\nu+l_\pi} \hat{S}
\hat{K} \hat{j}_\pi \hat{j}_\nu \hat{l}_\pi \hat{l}_\nu \langle
l_\pi l_\nu 0 0 \lvert L 0 \rangle \times \nonumber \\
&\begin{Bmatrix}
K & S & L \\
j_\pi & \frac{1}{2} & l_\pi \\
j_\nu & \frac{1}{2} & l_\nu
\end{Bmatrix},
\end{eqnarray}

\noindent
and we remind that $L$, $S$, and $K$ are the orbital, spin, and total
rank of the transition operators, respectively, and $l_\tau=k_\tau$ if
$k_\tau >0 $ and $l_\tau=\mid k_\tau\mid -1$ if $k_\tau <0 $.

Until now, the nucleon wave functions are expressed in a fully
relativistic framework, being the solutions of the Dirac equation.
However, within the nuclear shell model, the nucleon wave functions
are expressed as solutions of the single-particle Schr\"{o}dinger
equation, introducing the auxiliary harmonic-oscillator potential.

\noindent
Such an inconsistency impacts especially on the calculation of the
{\it relativistic} matrix elements, and this problem may be tackled in
two ways.
The first approach is to resort to a non-relativistic reduction of the
Dirac equation by considering the non-relativistic limit of the
coupled Eqs. (\ref{dirac1},\ref{dirac2}), namely the kinetic energy
$T$ and the auxiliary potential $V(r)$ satisfy the conditions $T=E -
M_N \ll 2M_N$ and $V(r) \ll 2M_N$ \cite{Rose54}.
Within this limit, the function $g_\kappa$ becomes the solution of the
Schr\"{o}dinger equation, and $f_\kappa$ is related to $g_\kappa$
through the relation
\begin{equation}
\label{fg}
f_\kappa(r)= \frac{1}{2M_N}\left(\frac{d}{dr}+\frac{\kappa+1}{r}\right)g_\kappa(r)~.
\end{equation}

However, as it was observed in Ref. \cite{Behrens71}, whether or not
such a limit of the Dirac equation is a good approximation to
calculate the {\it relativistic} NMEs is a difficult question to
answer, since to test this approach a fully relativistic calculation
should be performed and compared with the approximated results.

Moreover, it should be noted that, if we consider the
relation (\ref{fg}), the radial function $f_\kappa(r)$ is suppressed
by a factor $1/(2M_N)$ with respect to the function $g_\kappa(r)$, but
the {\it relativistic} form factor, even if it is small with respect
to the {\it non-relativistic} ones, it has been found to be relevant
to determine both the shape of the energy spectrum and the half-life of
the \bb~ decay
\cite{Behrens93,Kirsebom19,Dag20,Kossert2022,Tc99arxiv}.

Its relevance, within such a reduction of the Dirac equation, is also
stressed in Ref. \cite{Kostensalo21}, where a study of the dependence
of the energy spectrum and half-life of the fourth-forbidden \bb~
decay of \Cd~ with respect to the quenching factor $q$ of the axial
coupling constant $g_A$ has been carried out.
As a matter of fact, the authors show that a fitting procedure of both
$q$ and of the {\it relativistic} form factor is needed to reproduce the
experimental shape of the energy spectrum as well as the observed
half-life.

An alternative approach to calculate the {\it relativistic} NMEs, is
to resort to the conserved vector current theory (CVC) \cite{Behrens71},
which leads to derive a connection between the {\it relativistic} NMEs and the {\it
  non-relativistic} ones, that is developed as a relation between the
corresponding form factors.
An early application of this approach can be found in Ref. \cite{Behrens93}.

In particular, for the four decays under investigation, since we stop
at the leading order in the expansion of $F_{KLS}(q^2)$ ($N=0$ in
Eq. (\ref{Fexp})), the {\it relativistic} form factors entering in
Eq. (\ref{CW}) are only the $^{V}F_{211}$, for the second-forbidden
decay of \Nb, \Tc, and the $^{V}F_{431}$ for the fourth-forbidden
decay of \Cd, \In.
These form factors, using the CVC relation \cite{Behrens71,
  Behrens93},  depend on $^{V}F_{220}$ and $^{V}F_{440}$,
respectively, through the relations
\begin{equation}
\label{EQCVC}
^{V}F_{211}=-\frac{1}{\sqrt{10}}RE_{\gamma}~^{V}F_{220}~,\quad
^{V}F_{431}=-\frac{1}{\sqrt{36}}RE_{\gamma}~^{V}F_{440}~,
\end{equation}
\noindent
where 
\begin{equation}
E_{\gamma}=[W_0+\Delta E_C-(M_\nu-M_\pi)]~,
\end{equation}
is the energy of the analogue electromagnetic transition
\cite{Smith70},  $M_{\nu(\pi)}$ is the neutron (proton) mass, and
$\Delta E_C$ is the Coulomb displacement energy that can be evaluated
in different ways.
$\Delta E_C$ can be evaluated in different ways, here we have used the
results of the fit procedure as outlined in Ref. \cite{Antony97},
obtaining $\Delta E_C=11.99$, 12.38, 13.27, and 13.48 MeV for \Nb,
\Tc, \Cd, and \In~ decays, respectively.

It is important to stress that these CVC relations have been employed
in different studies, leading to a general improvement of the
calculated spectra
\cite{Behrens93,Kirsebom19,Dag20,Kossert2022,Tc99arxiv} and
half-lives.

However, we point out that these relations are valid in the full
nuclear Hilbert space of the single-particle configurations, and it is
difficult to evaluate the impact of resorting to a truncated model
space, but, on the other hand, it should be noted that our approach
relies on the derivation of effective Hamiltonians and operators by
way of the many-body theory to account for the configurations outside
the model space.

Nevertheless, the CVC relations for the relativistic form factors
represent a viable route to get an estimation of the {\it
  relativistic} matrix elements entering the calculation, and/or, to
understand the reliability of the non-relativistic reduction of the
{\it relativistic} matrix elements.

\subsection{Effective shell-model decay operators}\label{effopsec}
In this section, we sketch briefly the procedure to derive effective SM
decay operators \thetaeff~ by way of many-body perturbation theory.

As is well known, the diagonalization of the \heff~ does not produce
the true nuclear wave-functions, but their projections onto the chosen
model space $P$.
Then, any decay operator $\Theta$ should be renormalized
by taking into account for the neglected degrees of freedom
corresponding to the $Q$ subspace.

The derivation of effective SM operators within a perturbative
approach traces back to the pioneering period of employing $NN$
potentials in SM calculations
\cite{Mavromatis66,Mavromatis67,Federman69,Ellis77,Towner83,Towner87}.
We have followed the method that has been introduced by Suzuki and
Okamoto \cite{Suzuki95}, which allows a derivation of decay operators
\thetaeff~which is consistent with the one of \heff, as presented in
Sec. \ref{effham}.
This is based on the perturbative expansion of a vertex function \tbox~ --
analogously with the derivation of \heff~ in terms of the \qbox~ --,
whose details may be found in Refs. \cite{Suzuki95,Coraggio20c}.

According to such a procedure, the starting point is the perturbative
calculation of two energy-dependent vertex functions:

\[
\hat{\Theta} (\epsilon) = P \Theta P + P \Theta Q
\frac{1}{\epsilon - Q H Q} Q H_1 P ~, \]
\[ \hat{\Theta} (\epsilon_1 ; \epsilon_2) = P H_1 Q
\frac{1}{\epsilon_1 - Q H Q} Q \Theta Q \frac{1}{\epsilon_2 - Q H Q} Q H_1 P ~,\]

\noindent
and of their derivatives calculated in $\epsilon=\epsilon_0$,
$\epsilon_0$ being the eigenvalue of the degenerate unperturbed
Hamiltonian $H_0$:

\[
\hat{\Theta}_m = \frac {1}{m!} \frac {d^m \hat{\Theta}
 (\epsilon)}{d \epsilon^m} \biggl|_{\epsilon=\epsilon_0} ~, \]
\[ \hat{\Theta}_{mn} =  \frac {1}{m! n!} \frac{d^m}{d \epsilon_1^m}
\frac{d^n}{d \epsilon_2^n}  \hat{\Theta} (\epsilon_1 ;\epsilon_2)
\biggl|_{\epsilon_1= \epsilon_0, \epsilon_2  = \epsilon_0} ~\]

Then, a series of operators $\chi_n$ is calculated:

\begin{eqnarray}
\chi_0 &=& (\hat{\Theta}_0 + h.c.)+ \hat{\Theta}_{00}~~,  \label{chi0} \\
\chi_1 &=& (\hat{\Theta}_1\hat{Q} + h.c.) + (\hat{\Theta}_{01}\hat{Q}
+ h.c.) ~~, \nonumber \\
\chi_2 &=& (\hat{\Theta}_1\hat{Q}_1 \hat{Q}+ h.c.) +
(\hat{\Theta}_{2}\hat{Q}\hat{Q} + h.c.) + \nonumber \\
~ & ~ & (\hat{\Theta}_{02}\hat{Q}\hat{Q} + h.c.)+  \hat{Q}
\hat{\Theta}_{11} \hat{Q}~~. \label{chin} \\
&~~~& \cdots \nonumber
\end{eqnarray}

\noindent
At the end, \thetaeff~ is written in the following form:
\begin{equation}
\Theta_{\rm eff} = H_{\rm eff} \hat{Q}^{-1}  (\chi_0+ \chi_1 + \chi_2 +\cdots) ~,
\label{effopexp}
\end{equation}

\noindent
 the $\chi_n$ series being arrested in our calculations at
  $n=2$, and the $\hat{\Theta}$ function expanded up to third order in
  perturbation theory.

In Refs. \cite{Coraggio18,Coraggio19a,Coraggio20a} we have tackled
the issue of the convergence of the $\chi_n$ series and of the
perturbative expansion of the \tbox, showing the robustness of such a
procedure.

It is worth to point out that, even if the decay operator has a
one-body structure, the shell-model effective operator has many-body
components which account for a number of valence of nucleons larger
than one \cite{Ellis77,Coraggio20a}.
As a matter of fact, the fourth-forbidden \bb~ decay of $^{115}$In into
$^{115}$Sn involves 37 valence nucleons outside the doubly-magic
$^{78}$Ni, then for such a process \thetaeff~ should contain
contributions up to a 37-body term.

\begin{figure}[ht]
\includegraphics[scale=0.90,angle=0]{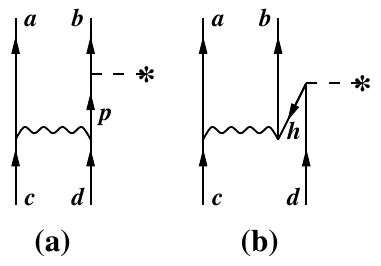}
\caption{Second-order two-body diagrams which are included in the
  perturbative expansion of \thetaeff.
  The dashed line indicates the bare second- and fourth-forbidden
  \bb-decay operators $\Theta$, the wavy line the two-body potential \vlwk. 
  For the sake of simplicity, for each topology only one of the diagrams
 which correspond to the permutation of the external lines has been drawn.}
\label{figeffop2b}
\end{figure}

The shell-model code KSHELL can employ transition operators with a
one- and two-body components \cite{KSHELL}, then for the calculation of
\bb-decay effective operators we include just the leading terms of these
many-body contributions in the perturbative expansion of the \tbox,
namely the second-order two-body diagrams (a) and (b), that are
reported in Fig. \ref{figeffop2b}.
  
The two topologies of second-order connected two-valence-nucleon
diagrams (a) and (b) accounts for the so-called ``blocking effect'',
which is necessary to consider the Pauli exclusion principle in systems with
more than one valence nucleon \cite{Towner87}, if the transition
operator has a one-body structure.
It is worth to point out that these two-body contributions to the
effective decay operators mirror the role of induced three-body forces
in the calculation of the \qbox, which we have introduced in section
\ref{effham}.

In the present work, the decay operators $\Theta$ are the one-body
electric-quadrupole $E2$ transition $q_{\scriptscriptstyle{p,n}}r^2
Y^2_m(\hat{r})$ -- the charge $q_{\scriptscriptstyle {p,n}}$ being $e$
for protons and 0 for neutrons --, as well as the one-body second- and
fourth-forbidden \bb-decay operators, as introduced in
Sec. \ref{operators}.

\section{Results}
\label{results}
In this section we present the results of our SM calculations.
First, we compare theoretical and experimental low-energy
spectroscopic properties of the parent and daughter nuclei under
investigation, namely \Nb, \Mo, \Tc, \Ru, \Cd, \Inn, \In~ and \Sn. 
Successively, we focus to study \bb-decay properties between the
ground states (g.s.) of these nuclei.

We point out that, rather than compare the experimental and calculated
half-lives, we consider the quantity \logft~ to describe the strength
of a $\beta$-transition. 
This is defined as the logarithm of 
\begin{equation}
 ft=\frac{\kappa}{\tilde{C}}\int^{w_0}_{1}p_ew_e(w_0-w_e)^2F(Z,w_e)dw_e ~.
\end{equation}

It is worth to stress again that all the calculations
have been performed without any empirical renormalization of \heff,
as well as without resorting to quenching factors of the axial
constant $g_A$.

\subsection{Spectroscopy}
\begin{table}[ht]  
\caption{Theoretical versus experimental low-energy levels of the
  nuclei under investigation.}
\begin{ruledtabular}
\begin{tabular}{ |c|c|c|c| }
\label{spec99}
&$J^\pi$  & E$^{th}$ (MeV) & E$^{Exp}$ (MeV) \\ 
\hline
\Nb&$6^+$  &  0.000 & 0.000 \\ 
&$3^+$  & 0.014    & 0.041 \\  
&$4^+$  & 0.014   & 0.058  \\
&$7^+$    & 0.017 & 0.079 \\  
\hline
\Mo&$0^+$ & 0.000 & 0.000 \\ 
&$2^+$&0.836    & 0.871 \\  
&$4^+$ &1.450   & 1.574 \\
\hline
\Tc&$\frac{9}{2}^+$  &  0.000 & 0.000 \\ 
&$\frac{7}{2}^+$  &   0.140 & 0.140 \\  
&$\frac{1}{2}^-$  &  0.794 & 0.143  \\
&$\frac{5}{2}^+$    & 0.812 & 0.181 \\  
\hline
\Ru&$\frac{5}{2}^+$    & 0.000 & 0.000 \\ 
&$\frac{3}{2}^+$    & 0.343 & 0.096 \\  
&$\frac{3}{2}^+$    & 0.433 & 0.321  \\
&$\frac{7}{2}^+$  & 0.588 & 0.341 \\ 
\hline
\Cd&$\frac{1}{2}^+$  & 0.000 & 0.000 \\ 
&$\frac{11}{2}^-$  & 0.069 & 0.263 \\  
&$\frac{3}{2}^+$   & 0.019 & 0.299  \\
&$\frac{5}{2}^+$  & 0.401 & 0.316  \\
\hline
\Inn&$\frac{9}{2}^+$  & 0.000 & 0.000 \\ 
&$\frac{1}{2}^-$    & 1.375 & 0.392 \\  
&$\frac{3}{2}^-$    & 1.613 & 0.646  \\
&$\frac{5}{2}^+$    & 1.270 & 1.024 \\
\hline
\In&$\frac{9}{2}^+$  & 0.000 & 0.000 \\ 
&$\frac{1}{2}^-$   & 1.196 & 0.336 \\  
&$\frac{3}{2}^-$   & 1.445 & 0.597  \\
&$\frac{3}{2}^+$   & 2.444 & 0.828 \\  
\hline
\Sn &$\frac{1}{2}^+$   & 0.068 & 0.000 \\ 
&$\frac{3}{2}^+$  & 0.000 & 0.497 \\  
&$\frac{7}{2}^+$    & 0.233 & 0.612 \\ 
&$\frac{11}{2}^-$ &  0.396 & 0.713 \\  
 \end{tabular}
 \end{ruledtabular}
\end{table} 

In Table \ref{spec99} we compare the low-energy spectra of the nuclei
under investigation.

The calculated energy levels are in reasonable agreement with the
corresponding experimental states with a few exceptions, while the
comparison between theory and experiment for the observed $B(E2)$
(Table \ref{trans}) is less satisfactory for some transitions. 
In fact, except for the \Tc, almost all the other calculated $B(E2)$s
underestimate the experimental ones.

\begin{table}[ht]
\caption{\label{trans} Theoretical versus experimental \cite{ensdf}
  low-energy $B(E2)$ transition strengths, in W.u., of the nuclei under
  investigation.}
\begin{ruledtabular}
\begin{tabular}{ |c|c|c|c| }
Nucleus& $J_i \rightarrow J_f$ & Theory & Experiment \\ 
\hline
\Mo& $2+ \rightarrow 0^+$ & 7.9 & 26 (4) \\
& $4^+ \rightarrow 2^+$  & 7.7   & 16.0 (4) \\
\hline
$^{99}$Tc& $\frac{7}{2}^+ \rightarrow \frac{9}{2}^+$ &  21 & 30 (19) \\
& $\frac{5}{2}^+ \rightarrow \frac{9}{2}^+$  &  10.3 & 15.1 (5) \\
\hline
\Ru& $\frac{3}{2}^+ \rightarrow \frac{5}{2}^+$  &  7.4 & 50.1 (10) \\
\hline
\Cd& $\frac{3}{2}^+ \rightarrow \frac{1}{2}^+$  &  2 & 20 (8) \\
& $\frac{5}{2}^+ \rightarrow \frac{1}{2}^+$  & 7.0& 0.372 (25) \\
\hline
\Inn&$\frac{5}{2}^+ \rightarrow \frac{9}{2}^+$ & 7.2 & 3.9 (4) \\
\hline
\In&$\frac{1}{2}^+ \rightarrow \frac{3}{2}^+$ &  17 & 121 (23) \\
\hline
\Sn& $\frac{3}{2}^+ \rightarrow \frac{1}{2}^+$ &  0.1 & 2.1 (6) \\
& $\frac{7}{2}^+ \rightarrow \frac{3}{2}^+$ &  0.010 & 0.130 (4) \\
 \end{tabular}
 \end{ruledtabular}
\end{table} 

Here, it is worth to notice that a sound shell-model description of
these nuclei is not an easy task for different reasons, especially in
a parameter free calculation.
In fact, we start from a $^{78}$Ni core, and, therefore, the number of
valence particle is sizeable, ranging from 16 to 37. 
Then, the inclusion of the contributions from three-body diagrams is
only the leading order of many-body contributions, whose role grows
when increasing the number of valence nucleons.
Besides this, Cd, In and Sn isotopes are at the limit of the proton
configuration space and, therefore, $Z=50$ cross-shell excitations may
play a relevant role in the renormalization of the electric-quadrupole
transition operator \cite{Coraggio15a,Coraggio16a}.
Actually, the enlargement of the proton model space, in order to
account explicitly for such excitations, is out of
 the present computational resources since the
dimensions of the Hamiltonians to be diagonalized could
reach $\approx 10^{13}$. 

\subsection{Forbidden $\beta$ decays of \Nb, \Tc, \Cd, and \In}
\label{Sec:spectra}
We focus now on the properties of the \bb~decay between ground states,
and it is important to start by discussing the relevance of the CVC
relations (See Eq. \ref{EQCVC}) in determining the {\it relativistic}
form factors.

As shown in Table \ref{TAB:FormF}, in the case of \Nb, \Tc~
second-forbidden \bb~ decays, the {\it relativistic} form factors
$^{V}F_{211}$ obtained using the bare operator is zero since the SPMEs
of the corresponding operator (Eq. \ref{rSPME}) are identically zero
in the model space.
Even though the renormalization procedure gives SPMEs different from
zero, the value of the $^{V}F_{211}$ form factors calculated using the
effective operator has an opposite sign, and it is a factor two
smaller than the form factor obtained by using the CVC relation
($^{V}F_{211}^{CVC}$).

\begin{table}[ht]
\caption{\Nb, \Tc, \Cd, and \In~ \bb~ decay {\it relativistic} form factors
  determined with and without resorting to the CVC relations, and the
  {\it non-relativistic} form factors connected with the {\it
    relativistic} ones by CVC. The values are in adimensional units.}
\label{TAB:FormF}
\begin{ruledtabular}
\begin{tabular}{ |c|c|c|  }
\Nb& Bare&Effective \\ 
\hline
$^{V}F_{211}$&0.000  &0.009    \\
$^{V}F_{211}^{CVC}$&-0.031& -0.016 \\
$^{V}F_{220}$&0.304  &0.164   \\
\hline
\Tc& Bare&Effective \\ 
\hline
$^{V}F_{211}$&0.000  &0.008    \\
$^{V}F_{211}^{CVC}$&-0.030 & -0.017  \\
$^{V}F_{220}$&0.286  &0.161   \\
\hline
\Cd& Bare&Effective \\ 
\hline
$^{V}F_{431}$& 0.0003& -0.008 \\
$^{V}F_{431}^{CVC}$&0.032&0.015  \\
$^{V}F_{440}$&-0.521  &-0.237 \\
\hline
\In& Bare&Effective \\ 
\hline
$^{V}F_{431}$&-0.0004 & -0.009\\
$^{V}F_{431}^{CVC}$& 0.031&0.017 \\
$^{V}F_{440}$&-0.473  &-0.267 \\
 \end{tabular}
\end{ruledtabular} 
\end{table}   

As regards the fourth-forbidden form factors of \Cd, it is interesting
to note that, in this case, the bare value of the {\it relativistic}
form factor $^{V}F_{431}$ has the same sign of $^{V}F_{431}^{CVC}$,
but it is two order of magnitude smaller. 
The effect of the renormalization is remarkable, but, as it happens
for the {\it relativistic} form factor of the \Nb, \Tc~ decays, the
final result again has an opposite sign and is a factor two smaller
with respect to the one calculated with the form factor
$^{V}F_{431}^{CVC}$.
 
The same considerations may be drawn for \In, except that, using the
bare operator, the sign of $^{V}F_{431}$ form factor is not consistent
with the one obtained with the $^{V}F_{431}^{CVC}$ one.

These results point out to a problem in determining the {\it
  relativistic} form factors within the non-relativistic reduction of
the Dirac equation for the nuclei under investigations. 
On the above grounds, in the following calculation of the \bb~ decay
properties we use the CVC relations for the {\it relativistic} form
factors.

\begin{table}[h]
\caption{\label{logft} Theoretical and experimental \logft~
  values. Data are taken from Ref. \cite{ensdf}.}
\begin{ruledtabular}
\begin{tabular}{|c|c|c|c|}
&Bare &Effective& Exp. \\ 
\hline
\Nb &11.30& 11.58&11.95 (7)   \\ 
\Tc &11.580& 11.876& 12.325 (12) \\ 
\Cd  & 21.902& 22.493& 23.127 (14) \\  
\In &21.22& 21.64& 22.53 (3) \\
 \end{tabular}
\end{ruledtabular} 
\end{table} 

We start the discussion of our results from the comparison between our
calculated values of the \logfts~ and the experimental ones, as they
are reported in Table \ref{logft}.
There, we have reported the \logfts~ obtained using both the bare
second- and fourth-forbidden \bb-decay operators and the effective
ones.
It is worth stressing again that, as we have reported in
Sec. \ref{effopsec}, our SM effective operators consists of one- and
two-body components.

As can be seen, the results that are obtained by employing the bare
\bb-decay operators underestimate the experimental \logfts, a result
that is consistent with the general consideration that nuclear models,
which operate truncation of the Hilbert space of the single-nucleon
configurations, require the introduction of a quenching factor $q$ in
order to reproduce the experimental half-lives for allowed \bb-decay
transitions (see for example Refs. \cite{Martinez-Pinedo96,Barea15}).

On the other hand, the calculations employing the SM effective
operators provide results that substantially recover the gap with
respect to the experimental \logfts, a result that is consistent with
our previous studies of the allowed \bb-decay within the realistic
shell model \cite{Coraggio17a,Coraggio19a,Coraggio22a,Coraggio24a}.

In order to discuss the role of the renormalization of the \bb-decay
operator on the calculation of the shape of forbidden \bb-decay energy
spectra, we evaluate the quenching factors that are needed to tune the
axial coupling constant $g_A$ in order to obtain the same results we
have obtained for the \logfts~ by employing the SM effective
operators.
Using Eq. (\ref{intCeq}), we obtain that the quenching factors that
reproduce the values in column ``Effective'' in Table \ref{logft} are
$q=0.27$, 0,50, 0.22, and 0.41 for \Nb, \Tc, \Cd, and \In~ decays,
respectively.

\begin{figure*}[ht]
\includegraphics[scale=0.35]{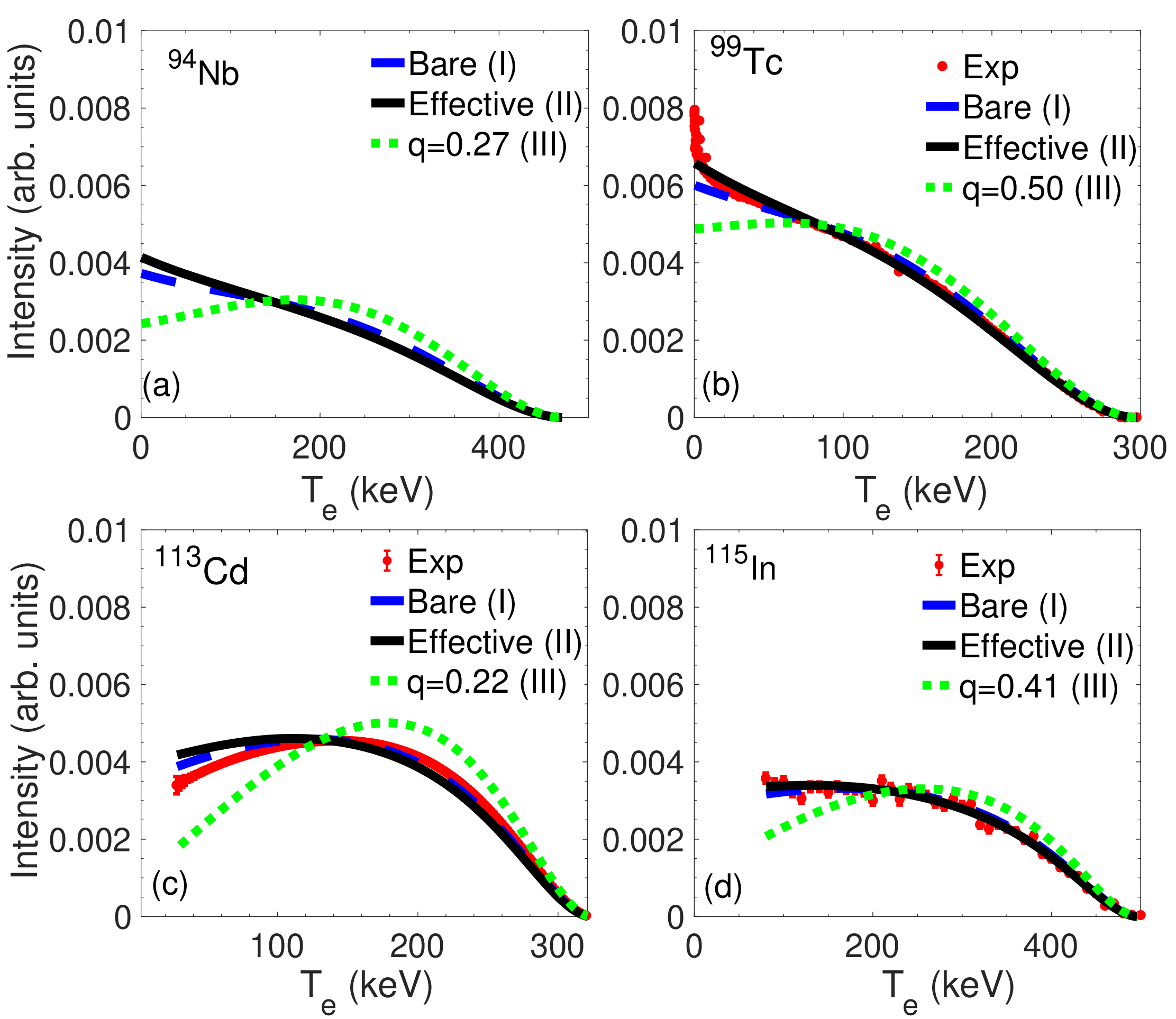}
\caption{\label{spectra} Theoretical and experimental normalized \bb-spectra
  of \Nb~ (a), \Tc~ (b),\Cd~ (c) and \In~ (d) as a function of the electron
  kinetic energy $\rm T_e$. The theoretical spectra are calculated
  with the bare operator (blue dashed line), the SM effective operator
  (continuous black line), and using the quenching factors for $g_A$
  extracted from Eq. (\ref{cva}) (green dashed line) (see text for
  details). The red dots corresponds to experimental values, where
  they are available \cite{Tc99arxiv,Belli07,Belli19,Pagnanini23}.}
\end{figure*}

In Fig. \ref{spectra}, the calculated and experimental normalized
spectra of the second-forbidden and fourth-forbidden decays, that are
under our present investigation, are reported.
The available data are drawn, with the corresponding errors, using red
dots.
It should be noted that the energy spectra are normalized in the
energy region of the available data, and that those for \Tc~ decay have
been extracted from Fig. (5) in Ref. \cite{Tc99arxiv}.
Moreover, the experimental spectra of \Tc~ and \Cd~ are obtained after
an unfolding procedure which takes into account the detector response
function, as discussed in Ref. \cite{Belli07} and
Refs. \cite{Tc99arxiv, Paulsen20}, respectively.
Conversely, in the energy spectrum of \In~ the detector response is 
not decoupled, leading to a more difficult direct comparison. 
However, according to the results presented in
Refs. \cite{Pagnanini23,Pagnanini24arxiv}, the detector response
effects are expected to be small and, in any case, not significant for
our purposes.

As regards \Nb~ decay, there are no experimental results, at present,
and they are normalized in the full range of the kinetic-energy interval.

The calculated values are labelled and drawn as follows:
\begin{itemize}
\item[(I)] the calculated values, obtained using the bare operators, follow
  the dashed blue line;
\item[(II)] the spectra calculated using the SM effective operators are
  drawn with a continuous black line;
\item[(III)] finally, we report also the results coming out by using the bare
  operator, but quenching the axial coupling constant $g_A$ with the
  $q$ factors that reproduce the theoretical \logfts~using the
  effective decay operators.
\end{itemize}

From the inspection of Fig. \ref{spectra}, we can clearly assert that
the theoretical energy spectra, calculated employing the bare and the effective
\bb-decay operators, are in a very good agreement with the corresponding
experimental shapes \cite{Tc99arxiv,Belli07,Belli19,Pagnanini23}, for 
the forbidden \bb-decays of \Tc, \Cd, and \In. 
Moreover, all of them exhibit a small sensitivity to the renormalization of
the \bb-decay operators, with small differences appearing only at low
energy for all the decays ($\lessapprox 100$ keV). 

As regards the energy spectra obtained with the bare operator and
quenching $g_A$, we see that they show a pronounced disagreement with
the experimental shapes, for \Tc, \Cd, and \In, mostly in the low- and
intermediate-energy intervals.

\noindent
These results lead us to two main conclusions.

First, the calculated shapes of the normalized energy spectra are
substantially insensitive to the renormalization of the forbidden
\bb-decay operator by way of many-body perturbation theory, that is
our approach to realistic shell model, and are in a good agreement
with current data.

Second, it seems that the mere renormalization of the axial coupling
constant $g_A$ by a quenching factor $q$ makes it difficult to provide
simultaneously better \logfts~and shapes of the energy spectra which
reproduce the observed behavior.

It is worth mentioning that in
Refs. \cite{Haaranen16,Kostensalo17,Tc99arxiv} the authors carried out
a study of the sensitivity of their calculated spectra upon the
renormalization of the axial coupling constant, and they found a
noticeable dependence of their results on the choice of the quenching
factor $q$.

Now, in order to reach a better insight of the results, we analyze
the different components of the shape factor defined in
Eq. (\ref{intCeq}).
This factor can be divided in three components, the vector, axial and
vector-axial terms, namely
\begin{equation}
\label{cva}
    C(w_e)=C_V(w_e)+C_A(w_e)+C_{VA}(w_e),
\end{equation}
where $C_V$ contains the coupling constant $g^2_V$, $C_A$ contains
$g^2_A$ and $C_{VA}$ contains $g_V \cdot g_A$.

The integrated shape functions $\tilde{C}_k$ are reported in Table
\ref{IntC}, as well as the total value $\tilde{C}$ (see
Eq.(\ref{intCeq})) for all the decays under investigation.
As can be seen, at variance with what was observed in
Ref. \cite{Haaranen17}, for all the decays $\tilde{C}_{VA}$ is
positive and therefore it is summed in phase with the vector and axial
components.

Actually, such a result is a consequence of the CVC theory, since if
we do not use the CVC relations to determine the {\it relativistic}
form factor, the mixed terms $\tilde{C}_{VA}$ for the cases under
investigation become negative, and very close in absolute value to the
sum of the correspondent vector and axial components.
This is similar to the results reported in Ref. \cite{Haaranen17}, and
such a feature highlights the relevance of this form factor, as it
was also discussed in other papers
\cite{Behrens93,Kirsebom19,Dag20,Kossert2022,Tc99arxiv}.

Even though it is small compared to the other form factors, the {\it
  relativistic} one is relevant do determine $C(w_e)$, and, therefore,
shapes and half-lives, since, at variance with {\it non-relativistic}
form factors, it enters the quantity $M_K(k_e,k_\nu) $ of
Eq. (\ref{CW}) without any suppression coefficient (for the explicit
expression of $M_K(k_e,k_\nu)$ see Table 4 in Ref. \cite{Behrens71}).

The positive values of $\tilde{C}_{VA}$ explain the stability of the
shape with respect to the renormalization of the decay operators.
In fact, without using the CVC relations, as a consequence of the
delicate balance of the vector, axial and vector-axial terms, it is
obtained a shape of the energy spectrum that is very sensitive to the
renormalization procedure.

\begin{table*}[ht]
\caption{Integrated shape functions $\tilde{C}$ of the studied
  transitions and their vector $\tilde{C}_V$, axial-vector
  $\tilde{C}_A$, and mixed components $\tilde{C}_{VA}$.}
\label{IntC}
\begin{ruledtabular}
\begin{tabular}{ |c|c|c|c|c|c| }
Parent & Op & $\tilde{C}_V$ &$\tilde{C}_A$ &$\tilde{C}_{VA}$& $\tilde{C}$ \\ 
\hline
\Nb &Bare&$5.44\times 10^{-9}$&$1.23\times 10^{-8}$& $1.34\times 10^{-8}$&$3.11\times 10^{-8}$\\
      &Effective& $1.40\times 10^{-9}$&$ 9.07\times 10^{-9}$&$ 5.72\times 10^{-9}$&$1.62\times 10^{-8}$\\
\hline
\Tc &Bare&$3.10\times 10^{-9}$&$5.90\times 10^{-9}$& $7.15\times 10^{-9}$&$1.61\times 10^{-8}$\\
      &Effective& $8.82\times 10^{-10}$&$ 4.14\times 10^{-9}$&$ 3.14\times 10^{-9}$&$8.17\times 10^{-9}$\\
\hline
\Cd &Bare&$ 1.19\times 10^{-19}$&$3.72\times 10^{-19}$& $2.80\times 10^{-19}$ & $7.70\times 10^{-19}$\\
         &Effective&$2.14\times 10^{-20}$&$1.14\times 10^{-19}$&$6.20\times 10^{-20}$ &$ 1.98\times 10^{-19}$\\
\hline
\In &Bare&$6.14\times 10^{-19}$&$1.93\times 10^{-18}$&$1.15\times 10^{-18}$&$3.69\times 10^{-18}$\\
         &Effective&$1.75\times 10^{-19}$&$8.68\times 10^{-19}$ &$3.72\times 10^{-19}$&$1.42\times 10^{-18}$\\
\hline
\end{tabular}
\end{ruledtabular} 
\end{table*} 

\section{Conclusions and Outlook}
\label{conclusions}
This work is the first attempt to describe the features of forbidden
\bb-decays within the framework of the realistic shell model, without
resorting to any phenomenological quenching factor for the axial and
vector coupling constants.

Such a study represents not only a validation of our theoretical
framework to assess the reliability to predict $0\nu\beta\beta$
nuclear matrix elements \cite{Coraggio20a}, but can give also useful
information for the recent experimental studies of the electron energy
spectra of forbidden \bb-decays.

First, we have verified the ability of our effective Hamiltonian and
transition operators by comparing the calculated low-energy spectra
and $E2$ transition strengths of parent and daughter nuclei, involved
in the forbidden \bb-decays under consideration, with their
experimental counterparts.
The comparison of the spectroscopic data with the corresponding
experimental ones is quite satisfactory, especially if we consider the
large number of valence particles -- ranging from 16 to 37 --, which
characterizes the nuclear system we have investigated.

Then, we have calculated both the half-lives and the energy spectra of
the emitted electrons of the second-forbidden \bb-decay of \Nb~ and
\Tc, and the fourth-forbidden \bb-decay of \Cd~ and \In.

As regards the outcome of our calculation of the properties of the
forbidden \bb-decay processes under investigation, the results may be
outlined as follows:
\begin{itemize}
\item The exam of the theoretical \logfts~ and the experimental ones
  shows that the results that are obtained with bare operators always
  underestimate the data, a feature that is resembling the problem of
  the quenching of $g_A$ in the allowed \bb-decay transitions.
The theory moves towards experiment by employing the theoretical
effective operators, as expected.
\item Starting from the wave functions that are obtained through the
  diagonalization of our \heff, the shape of the calculated energy
  spectra is rather insensitive to the choice of the \bb-decay
  operator, bare or effective, and in both cases the reproduction of
  the observed normalized energy spectra is more than satisfactory.
\item The latter result seems to be unrelated to the considerations
  about the calculated \logfts, and the comparison with data.
In fact, using the bare operator, but introducing a quenching factor
of the axial constant to improve the reproduction of the experimental
\logfts, it results in a distortion of the shape of the energy
spectra, that affects the agreement with the observed ones.
\end{itemize}

\noindent
Wrapping up the results we have obtained, we may say that the goal to
obtain, on the same footing, a better reproduction of half-lives and
the shape of the energy spectrum of the emitted electrons in forbidden
\bb~decays, by employing effective decay operators, is a delicate matter.
As it has been mentioned in Sec. \ref{Sec:spectra}, such an issue was
met also in other studies \cite{Haaranen16,Kostensalo17,Tc99arxiv},
where the authors showed that, without a renormalization of the
\bb-decay operator that is framed in the many-body theory, the
reproduction of the observed properties of forbidden \bb~ decays
cannot rely only on the quenching of $g_A$, and other empirical
parameters should be considered.

It is worth pointing out that such an issue does not emerge in the
study of allowed \bb~ decays, since the calculated energy spectrum
does not depend on the nuclear matrix element of the electroweak
currents, and the most relevant observable to be tested is the
half-life.

Our considerations lead to suggest that the study of forbidden
\bb-decay processes could be a valuable tool to refine the theoretical
knowledge of the renormalization of transition operators, and to rule
out models that could be not reliable to predict the value of nuclear
matrix elements for decays, such as in the case of the \zbb~ decay.

On the above grounds, we plan to extend the present work by studying
forbidden \bb~ decays in other mass regions, close to nuclei that are
candidates for detecting the \zbb~ decay.
Another interesting subject could be to tackle the forbidden \bb-decay
problem starting from the derivation of electroweak currents by way of
the chiral perturbation theory, which represent the new frontier to
frame the nuclear many-body problem within its underlying fundamental
theory -- the QCD --, and it is currently employed to investigate GT
transitions with different nuclear approaches
\cite{Gysbers19,King20,Baroni21,Gnech21,Gnech22,King23,Coraggio24a}.

\section{Acknowledgments}
We thank J. Kostensalo, T. Miyagi, M. Biassoni, C. Brofferio,
S. Ghislandi,  A. Nava and L. Pagnanini for useful comments and
discussion.
G. D. G. acknowledges the support from the EU-FESR, PON Ricerca e
Innovazione 2014-2020 - DM 1062/2021.
We acknowledge the CINECA award under the ISCRA initiative and under
the INFN-CINECA agreement, for the availability of high performance
computing resources and support.
This work is partly supported by the Czech Science Foundation (Czech Republic) 
P203-23-06439S.
\bibliographystyle{apsrev}
\bibliography{paper_betaspectrum_1.bib}
\end{document}